\documentclass[showpacs,twocolumn,aps,prb]{revtex4-1}
\usepackage{amsmath}
\usepackage{amsfonts}
\usepackage{amssymb}
\usepackage{amsthm}
\usepackage{graphicx}
\usepackage{bbm}
\usepackage{bm}
\usepackage{bbold}
\newcommand{\sinc}{{\rm sinc}}

\begin{document}

\title{Laser induced modulation of the Landau level structure in  single-layer graphene}
\author{Alexander L\'{o}pez$^{1}$}
\email[To whom correspondence should be addressed. Electronic
address: ]{alexander.lopez@physik.uni-regensburg.de}
\author{Antonio Di Teodoro$^{1}$}
\author{John Schliemann$^{2}$}
 \author{Bertrand Berche$^{3}$}
\author{Benjamin Santos$^{4}$}
 \affiliation{1. School of Physics Yachay Tech, Yachay City of Knowledge 100119-Urcuqui, Ecuador\\
 2. Institute for Theoretical Physics, University of
Regensburg, D-93040
Regensburg, Germany\\
3. Statistical Physics Group,  
	Institut Jean Lamour\footnote{Laboratoire associ\'e au CNRS UMR 7198},
	CNRS -- Universit\'{e} de Lorraine - Campus de Nancy, B.P. 70239, 
	F - 54506 Vand{\oe}uvre l\`es Nancy Cedex, France, EU\\
4. INRS-EMT, Universit\'e du Qu\'ebec, 1650 Lionel-Boulet, Varennes, Qu\'ebec J3X 1S2, Canada
 }	
\date{\today}

\begin{abstract}
We present perturbative analytical results of the Landau level quasienergy spectrum, autocorrelation function and out of plane pseudospin 
polarization for a single graphene sheet subject to intense circularly polarized terahertz radiation. For the quasienergy 
spectrum, we find a striking non trivial level-dependent dynamically induced gap structure. This photoinduced modulation 
of the energy band structure gives rise to shifts of the revival times in the autocorrelation function and it also leads to modulation of the 
oscillations in the dynamical evolution of the out of plane pseudospin polarization, which measures the angular momentum 
transfer between light and graphene electrons. For a coherent state, chosen as an initial pseudospin configuration,
 the dynamics induces additional quantum revivals of the wave function that manifest as shifts of the maxima and 
minima of the autocorrelation function, with additional partial revivals and beating patterns. These additional maxima 
and beating patterns stem from the effective dynamical coupling of the static eigenstates. We discuss the possible experimental detection schemes of our 
theoretical results and their relevance in new practical implementation of radiation fields in graphene physics. 
 
\end{abstract}
\keywords{}
\pacs{78.67.Wj, 71.70.Di, 72.80.Vp}

\maketitle
\section{Introduction}
The dynamical control of the transport properties of Dirac fermions in the condensed matter realm is currently an 
intense research topic. These Dirac fermions have been shown to emerge as the 
low energy excitations of two-dimensional systems with a honeycomb lattice structure as it occurs in graphene \cite{novoselov1,geim,guinearmp}.  
Recent theoretical\cite{oka,foa,wu} 
and experimental\cite{ganichev} works have discussed the role of radiation fields in the manipulation of the transport 
properties of monolayer graphene samples. By focusing on the 
Terahertz frequency regime particular attention is paid to the tunability of the induced band gaps. In addition, 
the possibility of generating topological insulating 
behavior was theoretically put forward both in the static\cite{kane} and dynamical regimes.\cite{topological1,topological2,auerbach}

In presence of a perpendicular quantizing magnetic field ${\bf B}=B\hat{z}$, the static spectrum of single layer graphene 
posseses a $\sqrt{B}$ field dependence which strikingly contrasts the 
linear $B$ dependence for conventional non relativistic \textrm{2DEG}\cite{novoselov1}. In addition, the $n=0$ Landau level (LL) 
has only one sublattice occupied at each Dirac point. Considering the LL scenario and a topological contribution given by 
an excitonic gap the authors of reference \cite{rabill} predict the appearance of Rabi oscillations when the system's 
initial quantum state is prepared by means of a short electric pulse and the subsequent dynamics is controlled by the oscillations 
between the  dynamically coupled LL. 

In this work 
we theoretically analyze the dynamical manipulation of the LL structure of charge 
carriers on suspended monolayer graphene when a periodically driving radiation field 
is applied perpendicular to the sample. A similar set up was proposed in reference\cite{rusin}, 
where a Gaussian laser pulse is introduced via the dipolar approximation. In that work, 
the authors discuss the dynamics of {\it Zitterbewegung} which is described in terms of 
the radiation field emmited by the accelerated charge carriers in graphene. In our proposal 
we consider a continuous laser field and thus make use of Floquet's theorem to recast the 
dynamics in an explicitly time-independent fashion but without need to resort 
to infinite-dimensional Fourier-mode expansion. Our approach has the advantage of providing 
an analytical description of the driven evolution of relevant physical quantities as the 
pseudospin polarization which is a measure of the angular momentum exchange among the charge 
carriers and the radiation field\cite{regan}. In the following we explicitly show how our semi 
analytical results are relevant at high values of the quantizing magnetic field $B$, where the 
coupling to the radiation field leads to non trivial qualitative modifications of the dynamical 
behavior of relevant physical quantities within the perturbative regime. To the best of our knowledge,
our work constitutes the first approach to the photoinduced modulation of the Landau level structure
in single-layer graphene in presence of an intense and continuously applied laser field. However,
we would like to mention that a recent work\cite{intensefield} has addressed the role of intense radiation field.
Yet, the authors of this work consider a quantized radiation field and do not address the Landau level
structure scenario in single-layer graphene.

\noindent The paper is organized as follows. In section II we present the model and summarize the perturbative results for the quasienergy spectrum. In section 
III we study the dynamics of the autocorrelation function and pseudospin polarization for two initially prepared states. First we consider
an eigenstate of the static Hamiltonian that has vanishing pseudospin polarization in the static regime. Next, we present the 
same analysis for an initially prepared coherent state, with static finite pseudospin polarization 
and highlight the main differences with respect to the other initial configuration. 
In section IV we discuss our main results and in section V we present concluding remarks and argue on the experimental implementation of our proposed 
theoretical setup. Finally, in the appendix we summarize some mathematical calculations arising during the perturbative analysis.
\section{Model}\label{sec1}
In this section we focus on the low energy properties of non interacting spinless charge carriers in a suspended monolayer graphene subject 
to a perpendicular, uniform and constant magnetic field ${\bf B}=B\hat{z}$. The dynamics is governed by 
Dirac's Hamiltonian. In coordinate representation it reads
\begin{equation}
\mathcal{H}_\eta({\bf r})=v_{F}(\eta\pi_x\sigma_x+\pi_y\sigma_y),\label{dirac}
\end{equation}
where $v_{F}\sim 10^6m/s$ is the Fermi velocity in graphene. 
In addition, the canonical momenta  $\pi_j=p_j+eA_j$ ($j=x,y$) contain the vector potential 
($\nabla\times{\bf A}={\bf B}$), $-e$ is the electronic charge ($e>0$), and $\eta=\pm1$ describes the valley degree of freedom. 
Using the definition of the magnetic length $l_B^{-2}=e B/\hbar$ and the cyclotron energy $\hbar\omega_c=\sqrt{2} v_F\hbar/l_B$  
the Hamiltonian Eq.~(\ref{dirac}) at each K (K') Dirac point, which corresponds to $\eta=+1$ ($\eta=-1$), can be written 
in the form 
\begin{eqnarray}\label{h00}
H_{+1}=\hbar\omega_c\left(
\begin{array}{cc}
 0& a\\
 a^\dagger& 0 
\end{array}
\right)\\
H_{-1}=-\hbar\omega_c\left(
\begin{array}{cc}
 0& a^\dagger\\
a & 0 
\end{array}
\right)
\end{eqnarray}

\noindent where the annihilation and creation operators are defined by standard relations as 
\begin{eqnarray*}
 a=\frac{l_B(\pi_x-i\pi_y)}{\sqrt{2}}\qquad \textrm{and}\,&&a^\dagger=\frac{l_B(\pi_x+i\pi_y)}{\sqrt{2}}.
\end{eqnarray*}
The eigenenergies of the Hamiltonian (\ref{dirac}) are then 
\begin{equation}\label{energy0}
E^{s,\eta}_{n}=s\eta\sqrt{n}\hbar\omega_c
\end{equation}
with $s=\pm1$. Positive (negative) values of $s\eta$ represents the conduction (valence) band at each Dirac point. In addition, the integer quantum 
number $n=0,1,2\dots$ 
labels the  Landau level (LL) structure of monolayer graphene. Using the eigenstates $|n\rangle$ of the operator $a^\dagger a$, 
the corresponding eigenstates $|\varphi^{s,\eta}_{n}\rangle$ read
\begin{eqnarray}\label{ev0}
|\varphi^{s,+1}_{n}\rangle=\frac{1}{\sqrt{2}}\left(
\begin{array}{c}
 s|n-1\rangle\\
 |n\rangle
\end{array}
\right)\\
|\varphi^{s,-1}_{n}\rangle=\frac{1}{\sqrt{2}}\left(
\begin{array}{c}
 -s|n\rangle\\
 |n-1\rangle
\end{array}
\right)
\end{eqnarray}
for $n\neq 0$. The zero energy eigenstate ($n=0$) is given in each case by
\begin{eqnarray}\label{ev00}
|\varphi^{+1}_{0}\rangle=\left(
\begin{array}{c}
 0\\
 |0\rangle
\end{array}
\right)\\
|\varphi^{-1}_{0}\rangle=\left(
\begin{array}{c}
 |0\rangle\\
 0
\end{array}
\right).
\end{eqnarray}
Due to time-reversal symmetry, we have $\mathcal{T}H_{+1}\mathcal{T}=H_{-1}$.
Let us now consider the effect of intense circularly polarized terahertz electromagnetic radiation, incident perpendicularly to the 
sample. We assume that the beam radiation spot is large enough compared to the lattice spacing so we can neglect any spatial variation.
According to the standard light-matter interaction formulation, 
the dynamical effects of a monochromatic radiation field incident perpendicular to the sample can be described by means of a 
time-dependent vector potential
\begin{equation}
{\bf A}(t)=\frac{\mathcal{E}}{\omega}\left(\cos\omega t,\delta\sin\omega t\right),
\end{equation}
where $\mathcal{E}$ and $\omega$ are respectively the amplitude and frequency of the electric field given in turn by the standard relation
$\bm{\mathcal{E}}(t)=-\partial_t {\bf A}(t)$. In addition, we are using $\delta=+1$ ($\delta=-1$) for right (left) circular polarization.
We are using circular polarization because it has been shown to provide the maximal photoinduced bandgap\cite{dora}.
Starting from the ordinary dipolar interaction term $-e{\bf p}\cdot{\bf A}(t)$, introduced to the Tight-Binding Hamiltonian via the Peierls 
substitution, we can evaluate the effects of the driving at each Dirac point as
\begin{equation}
V_\eta=ev_F[\eta\sigma_xA_x(t)+\delta\sigma_yA_y(t)], 
\end{equation}
which explicitly reads
\begin{equation}
V_\eta=\xi\eta(\sigma_x\cos\omega t+\eta\delta\sigma_y\sin\omega t). 
\end{equation}
with the effective coupling constant $\xi=ev_F\mathcal{E}/\omega$. 
This makes the total Hamiltonian 
\begin{equation}\label{time}
H_\eta(t)=H_\eta+V_\eta(t),
\end{equation}
periodic in time $H_\eta(t+T)=H_\eta(t)$, with $T=2\pi/\omega$ the period of oscillation of the driving field. Therefore, if we focus
on the K Dirac point ($\eta=1$), the physics at the K' Dirac point ($\eta=-1$) can be easily found by the substitutions $\xi\rightarrow-\xi$
and $\omega\rightarrow-\omega$. 

\noindent Thus, let us focus on the K point physics and afterwards, we can make the necessary substitutions. In order to simplify the notation we set
$H_{+1}=H_0$ and $V_{+1}(t)=V(t)$. Hence, defining rising $\sigma_+$ and lowering $\sigma_-$ pseudospin operators by the standard formulas
\begin{equation*}
\sigma_\pm=\frac{\sigma_x\pm i\sigma_y}{2},
\end{equation*}
the time-dependent interaction potential can be rewritten as
\begin{equation} \label{potential}
V(t)=\xi(e^{-i\delta\omega t}\sigma_++e^{i\delta\omega t}\sigma_-),
\end{equation}

\begin{figure*}[ht]
\includegraphics[width=0.9\textwidth]{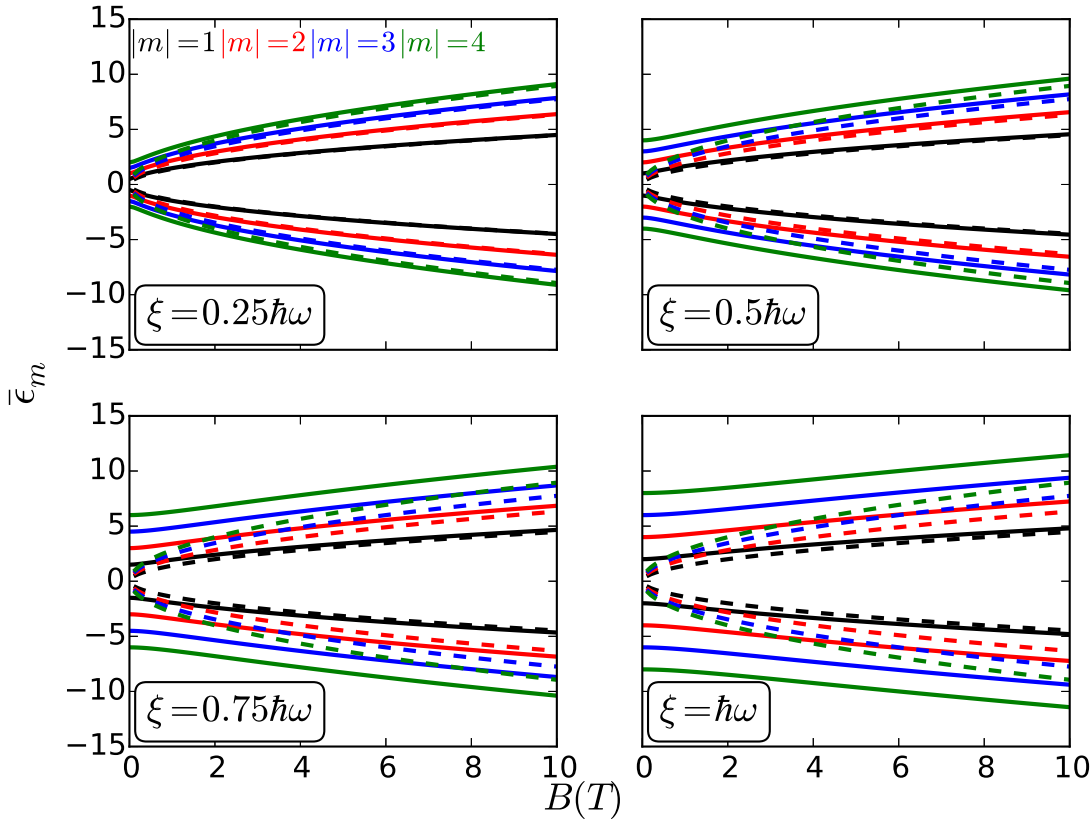} 
\caption{\label{fig:figure1}(Color online) Approximate mean energies for the driven scenario (continuous lines) as function 
of the quantizing magnetic field $B.$ The dotted lines represent the undriven spectra for the corresponding LL. We notice a 
level-dependent energy gapped that leads to non trivial behavior of physical quantities, as discussed below (see main text). 
}
\end{figure*}
Now we invoke Floquet's theorem which states that the time evolution operator of the system induced by a periodic Hamiltonian can be written in the form\cite{milena} 
\begin{equation}\label{unifull}
U(t)=P(t)e^{-iH_F t/\hbar}, 
\end{equation}with $P(t)$ a periodic unitary matrix and $H_F$ a time-independent dynamical generator referred to as the Floquet Hamiltonian. The eigenvalues of the Floquet Hamiltonian $H_{F}$ represent the quasienergy spectrum  of 
the periodically driven system. Typically, in order to solve for the quasienergy spectrum, one can expand each term of the time-dependent Schr\"odinger equation in 
Fourier space and numerically solve an infinite eigenvalue problem. Instead, we will take a perturbative approach as discussed below.\\

\noindent Accordingly, for our problem we can find approximate solutions to the dynamics by modifying slightly the analytical strategy presented 
in\cite{tilted}. Then, one finds that the excitation number operator $N_a$, defined as
\begin{equation}\label{nz}
N_a=\Big(a^\dagger a+\frac{1}{2}\Big)\mathbbm{1}+\frac{\sigma_z}{2},
\end{equation} 
which commutes with the Hamiltonian $H_0$ and satisfies the eigenvalue equation
\begin{equation}\label{excitations}
N_a|\varphi^s_{n}\rangle=n|\varphi^s_{n}\rangle. 
\end{equation}
$N_a$ generates a time-dependent unitary transformation 
$|\Psi(t)\rangle=P(t)|\Phi(t)\rangle$
given as
\begin{equation}
P(t)=\exp(-iN_a\delta\omega t),
\end{equation}
such that  the time-dependent Schr\"{o}dinger equation 
\begin{equation}\label{evolution}
i\hbar\partial_t|\Psi(t)\rangle=H(t)|\Psi(t)\rangle
\end{equation}
can be transformed with a time-independent operator $H_F$ governing the dynamics of the problem
\begin{equation}\label{effec1}
i\hbar\partial_t|\Phi(t)\rangle=H_{F}|\Phi(t)\rangle,
\end{equation}
where $H_F$ and $|\Phi(t)\rangle$ are the Floquet Hamiltonian and Floquet eigenstate, respectively. Doing the explicit calculation, 
$H_F$ is found to be given by
\begin{equation}\label{heff}
H_{F}=H_0-N_a\delta\hbar\omega+\xi\sigma_x.
\end{equation}

In the following we focus on recent experiments in the far infrared frequency domain\cite{ganichev} for which 
$\hbar\omega\approx10$ meV and we  consider values of the 
electric field intensities $\mathcal{E}\sim 0.15$ MV/m. Then one gets for the coupling constant 
$\xi\approx 10$ meV which, for frequencies $\omega$ in the terahertz domain leads to $\xi\approx\hbar\omega$. 
This value is an order of magnitude smaller than the Landau level separation $\hbar\omega_c\approx 116$ meV, for $B= 10$ T. 
For larger frequencies and stronger magnetic field intensities, the ratio $\xi/\hbar\omega_c$ tends to be smaller. 
Therefore, we can perform a perturbative treatment in the effective coupling parameter $\lambda=\xi/\hbar\omega_c<1$.\\ 
\noindent We should remark that although our radiation field is intense,
it is one order of magnitude smaller than the numerical estimates used in reference\cite{wu} for which one gets
$\mathcal{E}\sim 1.5$ MV/m. Yet, the experimental setup used in reference \cite{ganichev} consisted of infrared radiation field
with power $P=20$ mW, sample areas equal to $A=3\times 3\text{mm}^2$ and $A=5\times 5\text{mm}^2$; thus, one gets electric field
intensities of order $\mathcal{E}\sim 1$ kV/m, which in turn leads to $\xi\sim3$ meV. In this manner, our perturbative results 
would allow for an analytical treatment of future experimental extensions of the work described in \cite{ganichev}, in case 
they would include a quantizing magnetic field in their study. It would also allow for larger values of the radiation field 
intensity with novel photoinduced features in the Landau level structure of single- layer graphene as it is described in the 
following.

For this purpose we transform the Hamiltonian (\ref{heff}) as $H=e^{(\lambda/2) I_-}H_Fe^{-(\lambda/2) I_-}$, where we have introduced the antihermitian operator 
$I_-=a^\dagger\sigma_--a\sigma_+$. Evaluating up to first order we get 
\begin{equation}
H\approx H_F+\frac{\lambda}{2}[I_-,H_F]. 
\end{equation}
Evaluation of the commutator gives (in the appendix we summarize the explicit derivations)
\begin{eqnarray}
[I_-,H_F]=-\hbar\omega_c\Big[2N_a+\lambda(a^\dagger+a)\Big]\sigma_z,
\end{eqnarray}
and defining the shifted operator $b=a+\lambda$
one gets, to leading order in $\lambda$, the effective Hamiltonian
\begin{equation}\label{hfin}
H=\hbar\omega_c\Big(b^\dagger\sigma_-+b\sigma_+\Big)-\delta\hbar\omega N_b -\xi N_b\sigma_z,
\end{equation}
where we have introduced the shifted number operator
\begin{equation*}
N_b= b^\dagger b+\frac{\mathbb{1}+\sigma_z}{2}.
\end{equation*}
In equation (\ref{hfin}) we have neglected the additive higher order terms
\begin{equation}\label{pert1}
\Delta V= \hbar\omega\lambda(b^\dagger+ b)-\lambda^2\hbar\omega,
\end{equation}
which can be dealt with by higher order perturbation theory.
Thus,  the approximate quasienergies are  found to be given as
\begin{equation}\label{quasimono}
\epsilon^s_{m}=s\hbar\omega_c\sqrt{m}\sqrt{1+m\lambda^2},\,~\mod\hbar\omega,
\end{equation}
which can be rewritten as
\begin{equation}\label{quasimono2}
\epsilon^s_{m}=s\sqrt{m(\hbar\omega_c)^2+(m\xi)^2},\,~\mod\hbar\omega.
\end{equation}
Then, to this order of approximation, all quasienergies corresponding to the $m\neq0$ LL are shifted, whereas the $m=0$ 
remains insensitive to the radiation field. This shift of all but the $m=0$  quasienergy LL spectrum agrees with the 
result reported in reference \cite{bilayer} for 
bilayer graphene. Yet, a more detailed derivation by means of second order perturbation theory shows that there is a small 
$O(\lambda^4)$ energy correction due to first nondiagonal terms  in equation (\ref{pert1}) 
which couple all adjacent LL. This higher order corrections could be important at low quantizing magnetic fields for which the condition
$\lambda=\xi/\hbar\omega_c\approx1$ could be satisfied.\\ 
\noindent Let us now analyze some physical consequences of the radiation field on the Landau level structure of monolayer graphene with focus
on the interplay among the quantizing magnetic field and the light-matter interaction. To begin with, we notice that 
the quasienergies are defined up to multiples of $\omega$; therefore, a better physical characterization of the energy spectrum for the  driven  system
is provided by the mean energies\cite{milena}
\begin{equation}
\bar{\epsilon}^s_{m}=\epsilon^s_{m}-\omega \frac{\partial\epsilon^s_{m}}{\partial\omega}, 
\end{equation}
which are invariant under $\epsilon^s_{m}\rightarrow\epsilon^s_{m}+l\hbar\omega$, for $l$ being an integer. Doing the explicit calculation
the mean energies are found to be given by the expression
\begin{equation}
\bar{\epsilon}^s_{m}=s\Bigg(\frac{m(\hbar\omega_c)^2+2 m^2\xi^2}{\sqrt{m(\hbar\omega_c)^2+m^2\xi^2}}\Bigg), 
\end{equation}
where we remember the definition of the effective coupling to the radiation field as $\xi=ev_F\mathcal{E}/\omega$.\\
\noindent As can be seen in FIG. 1, these mean 
energies are plotted as function of the quantizing magnetic field $B$, for different values of the Landau level index changing the effective coupling $\xi$. 
We notice that, at intermediate light-coupling strength, the energy resolution of these levels becomes much better and could experimentally be tested for not so large quantizing
magnetic fields $B$. Moreover, we find that to this order of approximation the LL 
become gapped, with the striking feature that the photoinduced gap is level dependent. 
These gap openings appear except for the $m=0$ level which, as discussed before, remains insensitive to the radiation field.\\ 

Below we will deal with the photoinduced dynamical features and therefore, we give the corresponding normalized Floquet eigenstates 
for $m\neq0$
\begin{equation}\label{evfull}
|\psi^s_{m}\rangle=\left(
\begin{array}{c}
 sf^{-s}_{m}|m-1\rangle\\
  f^s_{m}|m\rangle
\end{array}
\right),
\end{equation}
where we have defined the coefficients
\begin{equation}\label{cs}
f^s_{m}=\sqrt{\frac{\epsilon_m+s m\xi}{2\epsilon_m}},
\end{equation}
with $\epsilon_{m}=|\epsilon^s_{m}|$. In addition, the zero energy  eigenstate ($m=0$) is still given by 
$\lvert\varphi_0\rangle$, i. e.,
\begin{equation}\label{ev0full}
|\psi_{0}\rangle=\left(
\begin{array}{c}
 0\\
 |0\rangle
\end{array}
\right).
\end{equation}
\begin{figure*}[ht]
\includegraphics[width=0.9\textwidth]{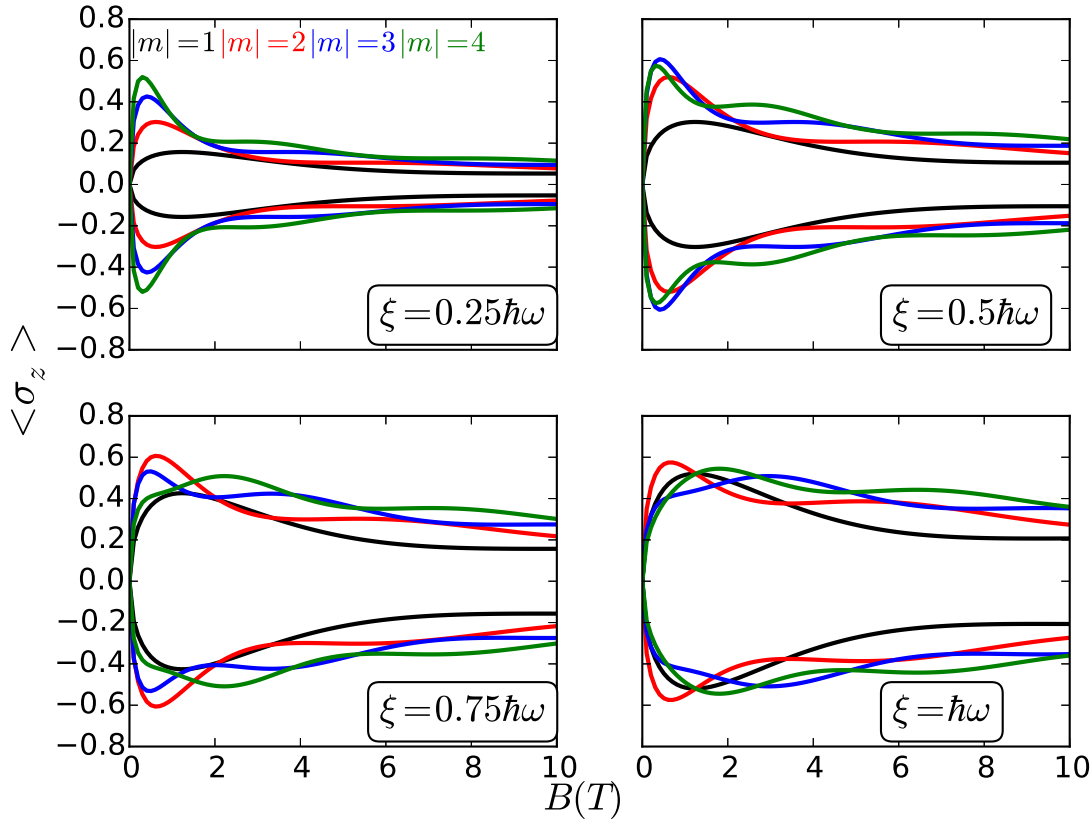} 
\caption{\label{fig:figure2}(Color online) Approximate mean pseudospin polarization $\langle\sigma_z\rangle$ for the driven 
scenario as given by equation (37), plotted as a function of the quantizing magnetic field $B.$ 
We have selected the first four LL $m=1,\dots4$, at
four values of the light-matter coupling strentgh. The lower panels show that at large magnetic fields ($B=10$T) both the $m=3$ and $m=4$ LL 
the pseudospin polarization still remains degenerate. Hence higher order perturbation terms should be necessary in order to lift this degeneracy. 
}
\end{figure*}
\section{pseudospin and autocorrelation function dynamics}
Now that we have found the approximate Floquet eigenstates and quasienergies, we explore other dynamical features 
of the driven LL configuration by evaluating the mean values of the pseudospin polarization operator and its relation to the autocorrelation function dynamics\cite{revivals}. 

\noindent For this purpose, let us first assume that the system
is initially prepared in an eigenstate of the Hamiltonian $H_0$
\begin{equation}\label{initialmono}
|\Psi(0)\rangle=|\varphi^s_{m}\rangle,
\end{equation}
with $m\neq0$.
In the Floquet basis (\ref{evfull}), the initial state is written as
\begin{equation}\label{initial}
|\varphi^s_{m}\rangle=\sum_{s'=\pm s}D^{ss'}_m|\psi^{s'}_m\rangle,
\end{equation}
where the expansion coefficients are given by
\begin{equation}
D^{ss'}_m=\frac{1}{\sqrt{2}}\left(f^{s'}_m+ss' f^{-s'}_m\right). 
\end{equation}

\noindent Taking into account that the $|\psi_m^{\pm}\rangle$ states are degenerate eigenstates of $N_b$, the unitary operator 
$e^{-iN_b \omega t}$ would just contribute a phase $e^{-i\omega m t}$.
 Using the Hamiltonian (\ref{hfin}) the evolved state can be written as
\begin{equation}
|\Psi(t)\rangle=\sum_{s'=\pm s}D^{ss'}_m(t)e^{-is'\epsilon_mt/\hbar}|\psi^{s'}_m\rangle,
\end{equation}
where we have introduced the time-dependent coefficients $D^{ss'}_m(t)=D^{ss'}_me^{-i\omega m t}$. 
\noindent Let us then consider the dynamics for the autocorrelation function and pseudospin polarization $\sigma_z(t)$  
operators from which we can respectively infer the feasibility of manipulating the polarization state of the sample and the dipole moment radiation 
emmited by the driven sample. It is important to remark at this point that studying the out of 
plane pseudospin polarization is a means of detecting the angular momentum exchange between the Dirac fermions in graphene and the 
circularly polarized radiation field, as can be inferred from the discussion in the recent literature about the role of 
$\sigma_z$ in describing the total angular momentum content of the system\cite{regan}.\\

\noindent With these ideas in mind, we first begin by evaluating the 
pseudospin polarization 
\begin{eqnarray}\label{szt}
\sigma_z(t,\xi)=\langle\Psi(t)|\sigma_z|\Psi(t)\rangle.
\end{eqnarray}
For the chosen initial state we find $\sigma_z(t,0)=0$.  When expression (\ref{szt}) is evaluated we find after some algebraic manipulations 
 
\begin{equation}\label{polafull}
\sigma_z(t,\xi)=\frac{2s\sqrt{m^3}\xi\hbar\omega_c}{\epsilon_m^2}\sin^2\epsilon_mt/\hbar.
\end{equation}
\noindent We note that for vanishing values of the coupling to the radiation field 
$\xi\rightarrow0$ one has $\sigma_z(t,0)=0$. Therefore, 
once the electromagnetic field is present the pseudospin oscillations are a manifestation of the angular momentum exchange among the radiation field and the charge carriers in graphene \cite{regan}. 
As discussed in the case of the quasienergies, we could better quantify the effects of the driving field 
by evaluating the average
\begin{equation}
\langle\sigma_z\rangle=\frac{1}{T}\int^T_0 dt\sigma_z(t,\xi), 
\end{equation}
with the period of the radiation field given as $T=2\pi/\omega$. 
Then we get the expression
\begin{equation}\label{polapro}
\langle\sigma_z\rangle=\frac{s\sqrt{m^3}\xi\hbar\omega_c}{\epsilon_m^2}\left(1-\sinc\,2\epsilon_mT/\hbar\right),
\end{equation}
with $\sinc\,{x}=\sin{x}/x$. \\

\noindent In figure  FIG. \ref{fig:figure2} we plot the behavior of $\langle\sigma_z\rangle$ as a function of the coupling strength $\xi$. 
At small $\xi=0.25\hbar\omega$ we see that the closer the state is to the $m=0$ LL the role of the radiation field in modifying its 
pseudospin polarization is less relevant and this is correlated to the fact that, within this regime, the gap openings seen in the mean 
quasienergy spectrum are not so noticeable at different values of the $m$ LL index. In addition, at large values of the  quantizing field, $B = 10T$, the value of the pseudospin polarization is almost the same at each value of the $m$ LL index. 
However, we see in the second upper panel that already at intermediate values of the relative coupling strentgh 
$\xi=0.5\hbar\omega$ that at large values of the quantizing field $B=10$T one can discern among the different LL polarization value
which serves to separate each level contribution to this quantity dynamical behavior. In the lower two panels we show that 
this physical picture becomes more obvious at larger values of $\xi$, with perfect separation for the contribution from the $m=1$ and $m=2$, levels. Yet,
the corresponding $m=3$ and $m=4$ contributions to the pseudospin polarization remain degenerate at large $B$ values. One would expect that this accidental crossings 
seen as a degeneracy would be removed by including higher order perturbative contributions, since then the coupling among nearby levels would lead to additional
splittings in the quasienergy spectrum that would in turn also lift this accidental pseudospin degeneracy.

In order to discuss more general dynamical features, one should consider a superposition state. However, in order to do so we must take into account
the feasibility of experimentally realizing such a superposition state. A paradigmatic case of such interesting superposition states is given by the coherent
state which are minimal uncertainty wave packets relevant for studying the classical states of the radiation field in the sense of being a classical counterpart of the quantum harmonic oscillator.
Since, as in the quantum harmonic oscillator, the Landau levels in graphene are eigenstates of the number operator, one should expect that these
coherent state superpositions would be interesting. Indeed, there has been recently some proposals to analyze the dynamics
of coherent electronic states in graphene nanomechanical resonators \cite{resonator}. By taking advantage of the intrinsic non linear nature of flexural modes in graphene, 
the authors of reference \cite{resonator} show that cat-like\cite{cat} states can be generated. We follow a different physical approach. Instead of resorting to nonlinearities of flexural modes 
we invoke the light-matter coupling as a mechanism for studying the evolution of an initially prepared coherent superposition state built from the Landau level
eigenstates described in section II and evaluate the induced pseudospin polarization effects in order to contrast to the results shown in FIG. \ref{fig:figure2}.
\begin{figure*}[ht]
\includegraphics[width=0.9\textwidth]{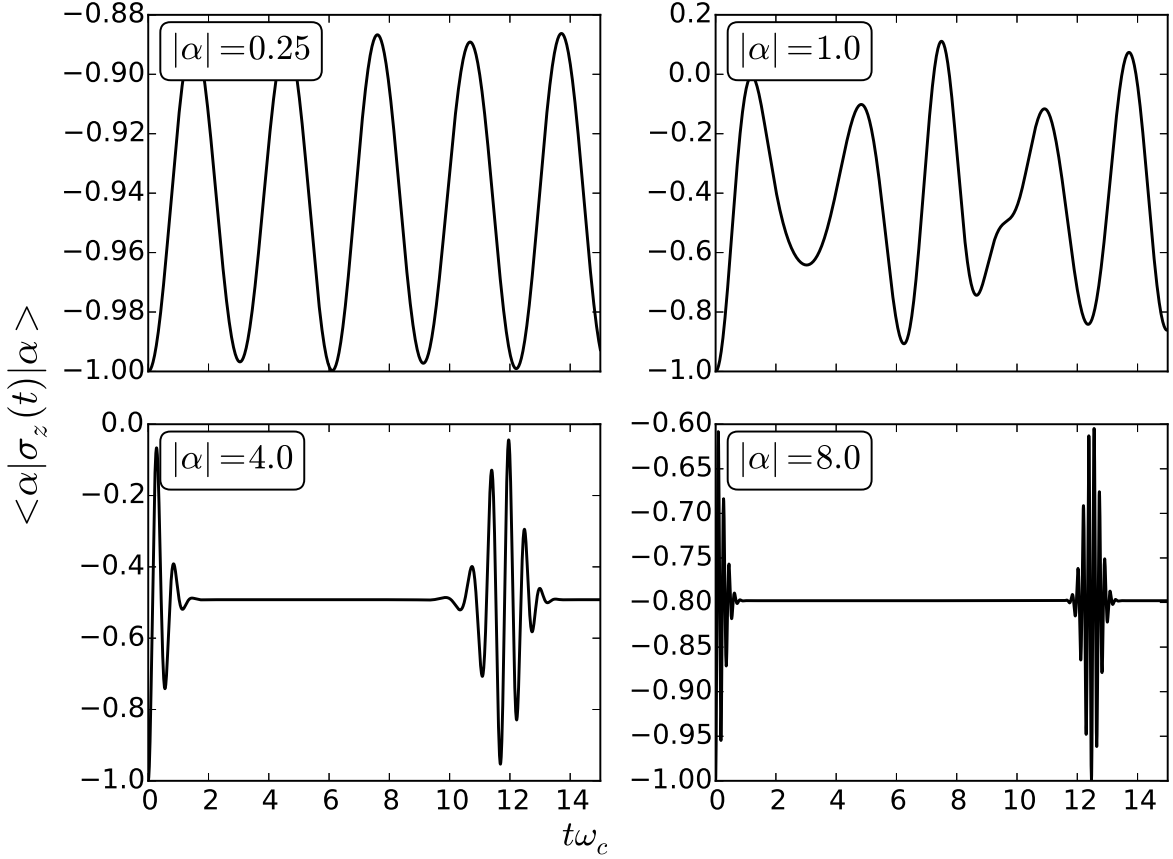}
 \caption{\label{fig:figure3} Time dependence of the pseudospin polarization  as given in equation 
 $(45)$. The panels show the dynamical behavior at four characteristic values $|\alpha|$. For the numerical evaluation of the series we have truncated at 
 $n=N=100$. The upper panels show the coherence among the lowest LL whereas the lower panel show  
 localization effects approaching the classical behavior, which corresponds to large values of $|\alpha|$. See the discussion in the main text.}
\end{figure*}

\noindent Formally speaking, the coherent state $|\alpha\rangle$ is defined by means of the eigenvalue equation

\begin{equation}\label{dqla}
A\lvert\alpha\rangle=\alpha\lvert\alpha\rangle, 
\end{equation}
where $A=a\mathbbm{1}$. Using the expansion

\begin{equation}
\lvert\alpha\rangle=c_0|\varphi_0\rangle+\bar{\sum_{sn}}c^s_{n}\lvert \varphi^s_{n}\rangle, 
\end{equation}
with $\bar{\sum}_{sn}$ representing a summation for all $n\neq0$. 
Assuming, without loss of generality, the symmetric scenario $c^{-s}_n=c^s_n$,
one finds that the coherent state is given as 
\begin{equation}
\lvert\alpha\rangle=e^{-\frac{\lvert\alpha\rvert^2}{2}}\Big(|\varphi_0\rangle+\frac{1}{\sqrt{2}}\bar{\sum_{sn}}\frac{\alpha^n}{\sqrt{n!}}\lvert \varphi^s_{n}\rangle\Big), 
\end{equation}
which can be shown to be normalized. If we now evaluate the mean value of the hermitian operator $A^\dagger A$ in the coherent state and use the definition
given in equation (\ref{dqla}), we get 
$\langle\alpha| A^\dagger A|\alpha\rangle=|\alpha|^2$. In addition, if we use the coherent state to evaluate the average of the number operator 
$N_a$ defined in equation (\ref{excitations}), it is not difficult to show that we also get 
$\langle\alpha|N_a|\alpha\rangle=|\alpha|^2$. Therefore, the coherent state parameter amplitude $|\alpha|$ is a measure of the mean number of Landau levels that are excited
and corresponds to the mean photon number in the context of quantum optics. Thus, we will use it as a control parameter to discuss
the properties of the physical quantities as follows.

\noindent In this case, we find for the pseudospin polarization 
\begin{equation}
\langle\alpha\lvert\sigma_z(t)\rvert\alpha\rangle=-e^{-\lvert\alpha\rvert^2}\sum_{n}\frac{\lvert\alpha\rvert^{2n}}{n!}
\Bigg(\frac{(\hbar\omega_c)^2\cos 2\epsilon_n t/\hbar+n\xi^2}{(\hbar\omega_c)^2+n\xi^2}\Bigg) 
\end{equation}
\begin{figure*}[ht]
\includegraphics[width=0.9\textwidth]{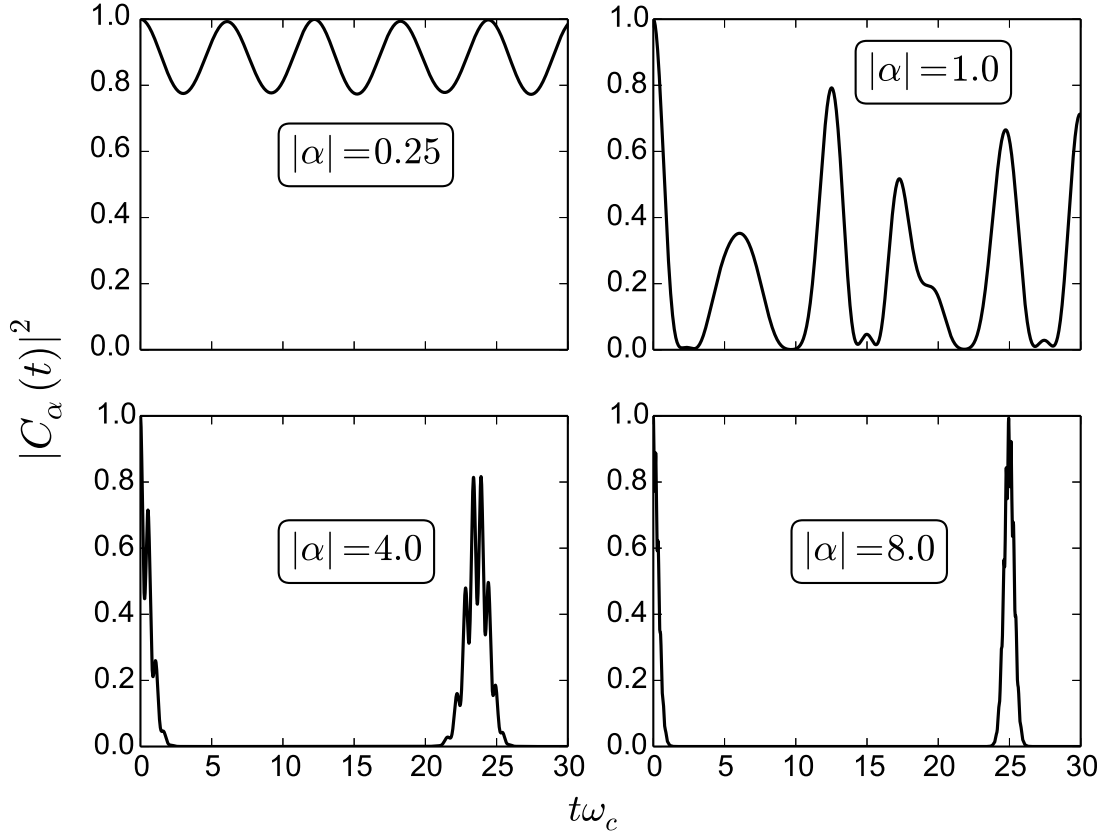}
 \caption{\label{fig:figure4-1} Time dependence of the autocorrelation function  as given in equation 
 $(48)$ truncated at $n=N=100$. For large $|\alpha|=4$ and $|\alpha|=8$,  it is the cyclotron frequency $\omega_c$ 
 instead of the driving frequency $\omega$ which determines the time scale for the revival times. Comparing to FIG. 3 it is apparent that information on the pseudospin dynamics
 can be indirectly inferred from the autocorrelation function dynamics. See discussion in the main text.}
\end{figure*}
This is plotted in FIG. 3 for $\xi=0.25\hbar\omega$ and four representative values of $|\alpha|$. In this figure we notice that for small values
of $|\alpha|$ the dynamics of the pseudospin polarization resembles the pattern for Rabi oscillations since the main contributions
would arise for the interference among the zero and first LL. Yet, no polarization inversion can be achieved within this regime.

However, once $|\alpha|=1$ the contribution from other LL states becomes increasingly important to the interference pattern
and the former Rabi oscillations become distorted. Moreover, at this value of $|\alpha|$, one can achieve the polarization 
inversion for large enough values of time in the long-term evolution. In the lower panels of FIG. 4 we find that for 
larger values ($|\alpha|=4$ and $|\alpha|=8$ ), we get a beating pattern showing a  
dynamical localization effect that is directly related to a collective behavior of the driven charge carries in graphene. 
We would like to remark that the problem of the population inversion has already been studied long time ago 
by Eberly et al\cite{eberly}. Their model corresponds to a two-level system coupled to a single mode radiation field 
(Jaynes Cummings Hamiltonian). In our approach, this in turn is given by the static graphene Hamiltonian for Landau levels 
written in equation (2). A larger value of the coherent state 
parameter $|\alpha|$ would imply larger mean Landau level occupation, and thus the second term in the numerator of the 
pseudospin polarization given in expression (43) would have a larger influence in the pseudospin inversion since larger n 
Landau levels would be occupied which will contribute a higher weight in the pseudospin polarization 
(see numerator in equation (20). Moreover, since the effective light-matter coupling strength $\xi$ also affects the 
phase, as given by the cosine term in equation (20) of the polarization, this interesting interplay forbids the inversion 
which is also a signature of the localization and beating effects shown on panels 3(c) and 3(d) and our results extend those 
found in the context of reference\cite{eberly}.

\noindent This physical picture for the coherent state dynamics can be complemented by studying the autocorrelation 
function $C_\alpha(t)=\langle\alpha|\Psi(t)\rangle$, which is found
to be given as
\begin{equation}
C_\alpha(t)=e^{-\lvert\alpha\rvert^2}\sum_{n}\frac{\lvert\alpha\rvert^{2n}}{n!}
\Bigg(\cos\epsilon_n t/\hbar+i\frac{n\xi}{\epsilon_n}\sin\epsilon_n t/\hbar\Bigg).
\end{equation}
The autocorrelation function provides additional physical information of the system since its Fourier
 transform is related to the local density of states\cite{revivals}.
Its time evolution is plotted in FIG. 4 choosing again an effective coupling strength value of $\xi=0.25\hbar\omega$ and the same values of $\alpha$ as in FIG. 3 showing the dynamical behavior of the pseudospin polarization. 
Comparing FIG.3 and FIG. 4 we see that partial revivals for the autocorrelation function are correlated to the beating or 
localization behavior of the pseudospin polarization. Therefore, one could indirectly gain information on the pseudospin dynamics by measuring the time revivals
\cite{revivals} 
which means detecting those times for which the wave packet reconstructs itself.
\begin{figure*}[ht]
\includegraphics[width=0.9\textwidth]{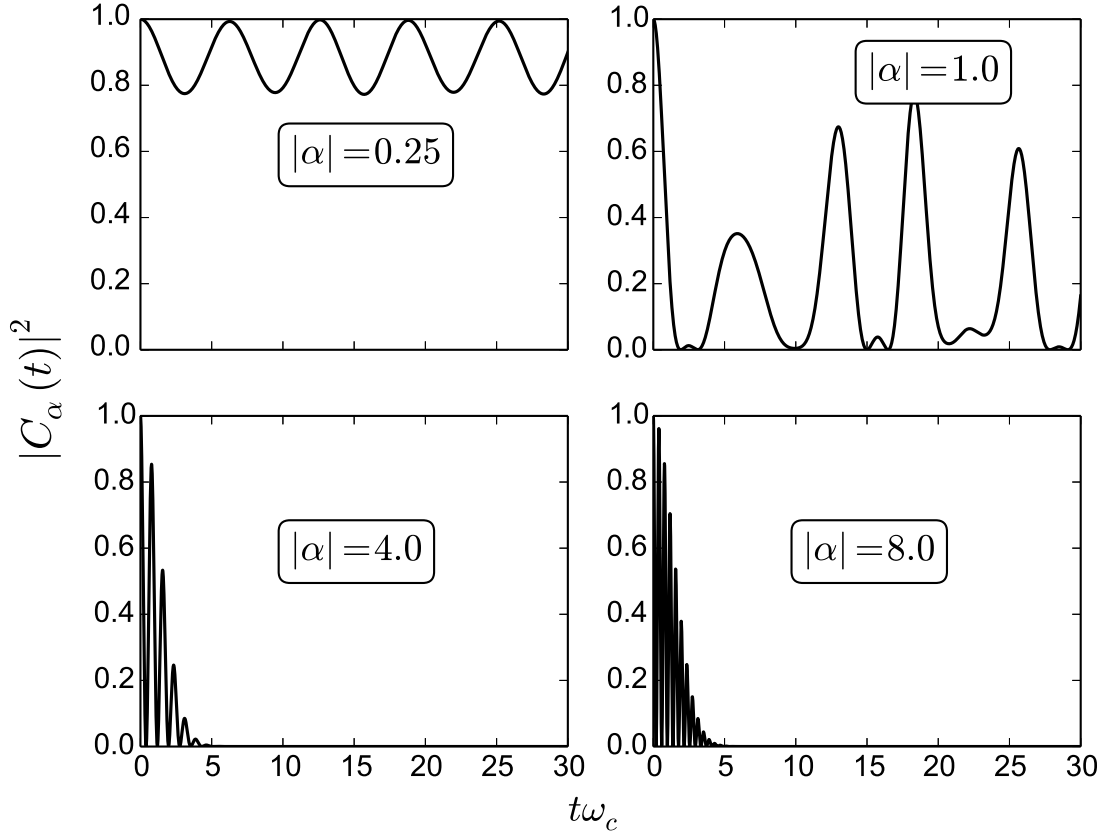}
 \caption{\label{fig:figure4-2} Time dependence of the autocorrelation function  as given in equation 
 $(48)$ truncated at $n=N=100$ in absence of radiation field $\xi=0$. In the upper panels we see that at small values of 
 $|\alpha|=0.25$ and $|\alpha|=1$, the static and driven 
 autocorrelation function are qualitatively similar. Yet, the  photoinduced quantum revivals at larger times are not seen in the two lower panels as compared to FIG.4 }
\end{figure*}
In order to show explicitly the role of the driving field we have plotted in FIG. 5 the static autocorrelation function. Comparing the lower right 
panels in FIG. 4 and FIG. 5, we find that the radiation field induces additional revivals in accordance to the beating pattern in the pseudospin oscillation, as seen in FIG.3.

\section{Discussion}
 We have shown that the radiadion field leads to a quasienergy spectrum with a level dependent gap, except for the 
  $m=0$ LL which, to leading order, remains insensitive to the radiation field effects. 
  In addition, we have found that a finite out of plane pseudospin polarization value can arise 
  for initial states that possess either a finite or vanishing initial value of $\sigma_z$. As we saw at the beginning of section 
  (\ref{sec1}) in presence of the radiation field, the total out of plane angular momentum component $j_z$ of the electrons is no longer a constant of motion. Therefore, the pseudospin oscillations are due to 
  the angular momentum exchange among the driving field and the charge carriers in single layer graphene.  For the coherent state a beating pattern emerges in the pseudospin polarization. In spite
  of the fact that the interaction is dictated by a periodic Hamiltonian we find that the dominant or characteristic time 
 scale for this collective behavior is the cyclotron frequency. We could expect this behavior of the wave packet at large values of 
 $|\alpha|$ as a measure of the classical behavior associated to the cyclotron problem but now realized with Dirac fermions
 in the LL quantized regime as was discussed by one of the authors in reference\cite{john}. We must remark that in order to be able to detect the reported effects an ensemble of 
 coherent states should be prepared each time, because the measurement process destroys the state. It should also be 
 remembered that the coherent state parameter $|\alpha|$ is determined
 by choosing appropiate values for the mean values of position and momentum in accordance to the following prescription
\begin{eqnarray}
\langle\alpha| x|\alpha\rangle&=&\ell_B\Re{(\alpha)},\label{x}\\
\langle\alpha| p_x|\alpha\rangle&=&\frac{1}{\ell_B}\Im{(\alpha)}\label{p}. 
\end{eqnarray}  
\noindent 
Another point to be highlighted is that for standard two dimensional electron gases the coherent state built from the 
Landau levels would remain coherent,  i.e., it will evolve in time in such a manner that
it would just oscillate in time around the prescribed mean values given in eqs. (\ref{x}) and (\ref{p}). More precisely, its Wigner function representation in phase space will
oscillate in time without deformation.
This is known to be a consequence of the fact that the dependence with the $n$ quantum number is linear. Yet, in  
graphene we have a $\sqrt{n}$ and this in turn prevents the coherent state to evolve in such a coherent manner. 

Therefore, the spreading of the wave packet and the corresponding appearance of the additional revival times at shorter time 
values is a direct consequence of the driving field that even at low coupling can induce interesting dynamical behavior as 
shown in FIG. 3 and FIG. 4. We supported this last statement by evaluating the static ($\xi=0$) autocorrelation function 
as shown in FIG. 5 and found that the second packet reconstruction (see lower panel for $|\alpha|=8$ in FIG. 4) around $t\omega_c=24$ 
is absent in the static regime.

We would also like to comment that  graphene subject to electromagnetic radiation without quantizing magnetic field has been discussed in several papers \cite{foa}. 
It has been shown that for zero momentum (k=0), the dynamical equation is exactly solvable and leads to a photoinduced (zero quantizing magnetic field) 
mass term that has opposite signs at the two valleys. Yet, at finite momentum the dynamics is no longer exactly solvable and one has to resort to numerical 
analysis by means of an infinite expansion in Fourier modes. This is why we consider our results to be a valuable tool in 
analytically describing the photoinduced pseudospin effects in the Landau level structure of monolayer graphene.

We now briefly mention that our values of quantizing magnetic fields are within experimentally accesible orders. For instance,
in the pioneering paper by Novoselov et al. in reference \cite{qhe}, the authors used a value of quantizing magnetic field 
$B=14$T to study the quantum Hall effect in graphene. Moreover, in a following classical paper, their results were extended 
up to values of $B=29$T and even $B=45$T for the quantizing magnetic field\cite{qhe2}. Here it is shown that at these values of magnetic field, the quantum Hall effect in graphene could be observed at room 
temperature that constitutes a breakthrough in the physics of quantum Hall phases. 

  \section{Conclusions}
  We have analyzed the dynamical modulation of physical quantities for Dirac fermions within the Landau level 
  quantized regime of single layer graphene subject to intense circularly polarized Terahertz radiation. To the best of our knowledge,
  this is the first analysis of the photoinduced manipulation of the LL structure in single-layer graphene subject to a
  continuously applied intense laser field instead of a pulsed one.
  By means of a perturbative analytical treatment we found very interesting physical features such  as a non trivial 
  level dependent dynamically induced gap structure.  Due to the angular momentum exchange among the radiation field and the charge carriers,  
  it also leads to modulation of the oscillations in the dynamics of the out of plane pseudospin polarization, even for superposition states with an 
  initially vanishing pseudospin value.  We also found that localization effects in the time evolution of the pseudospin polarization can be
  keep track of by measuring the revival times of a wave packet initially prepared as a coherent state. 
  The reported photoinduced gap modulation and pseudospin oscillations could be detected through the reemitted dipolar radiation 
  from the oscillating charge carriers as it was proposed in reference\cite{rusin}.\\
  
\noindent  We would also like to mention that for values of $\lambda\gg1$, one enters 
the so called ultrastrong coupling limit in the context of cavity quantum electrodynamics 
experiments. Within this parameter regime, it has been experimentally shown in reference 
\cite{ultrastrong} that explicit anti-crossings appear in the energy spectrum. This in turn corresponds to a breakdown of the 
Jaynes cummings approximation that describes resonant processes in the two-level problem coupled to a single mode radiation 
field. Our exact Floquet Hamiltonian given in equation (20) allows us to explore this 
ultrastrong coupling regime beyond the perturbative results presented and could be the subject of future 
work where one would expect non linear effects to be relevant in the graphene physics. For instance, as we 
already argued in the discussion section, one could expect that the accidental degeneracies seen in the
panels of FIG. 2, at large values of the quantizing magnetic field, would be lifted by a stronger coupling 
to the radiation field. 

\noindent Finally, we would like to mention that although we perform our analysis for neutral undoped graphene. Yet,
in the graphene literature the role of doping has been given special attention
since it might either modify the effective Fermi velocity of Dirac fermions\cite{doping0, doping1}. The first work\cite{doping0} 
shows that the energy spectrum near the charge neutrality point is non linear and no gap is found at energies even as close to 
the Dirac point as $~0.1$ meV. On the other hand, the authors of paper \cite{doping1} show that that electron interactions leave 
the graphene energy dispersion linear as a function of excitation energy even for energies within $\pm200$  meV of the Fermi energy.

\noindent Yet, it has recently been shown that doping effects might also lead to a bandgap opening in the graphene spectrum (see \cite{review-doping} and references therein).
In particular, the authors of the work\cite{review-doping} have discussed the tunability of the bandgap energy in single-layer graphene due to 
manganese oxide nanoparticles by means of an electrochemical method \cite{echemical}. This work reports a maximum value 
for the induced energy bandgap of $0.256$ eV. Although they do not discuss the quantized LL regime, we could incorporate 
these doping effects in our model by including a phenomenological diagonal term in the Landau level Hamiltonian given in 
equations (2) and (3). This mass-term would in turn be proportional to $\sigma_z$ and the valley index $\eta$, as required 
by time reversal invariance. Thus, one would expect an interesting interplay between the photoinduced gap openings and the 
static bandgap determined by doping effects. Yet, we considered undoped graphene in order to highlight the photoinduced 
LL dependent bandgap discussed above. 

\noindent Concerning the actual experimental observation of our predicted results, we consider 
that these could motive the exploration of larger intensities for the radiation field as given in the experimental setup in 
reference \cite{ganichev} when their model enters the Landau level regime. In this sense, we consider that our work would 
contribute to explore a new physical scenario within realistic parameter values to discuss the pseudospin physics in graphene, 
also taking into account the role of Zitterbewegung as it is discussed in reference\cite{rusin}, extending their results
to initially prepared coherent states when the radiation field is a monochromatic continuous laser field instead of a pulsed laser.\\ 

{\it{Acknowledgments}--} This work has been supported by Deutsche Forschungsgemeinschaft via GRK 1570 and Yachay Tech via 
the project ``Non dissipative transport in honeycomb lattice materials: Interplay of spin-orbit interaction, 
radiation fields and superconductivity''. AL thanks the University of Lorraine for 
partial financial support through research visits. BB thanks Yachay Tech for financial support for research visits.
\section{appendix}
\subsection{Perturbative calculation of effective Hamiltonian for single layer graphene}
In order to get the effective Hamiltonian for single layer graphene we need to evaluate the following expression
\begin{equation}
H=e^{\lambda/2 I_-}H_Fe^{-\lambda/2 I_-},
\end{equation}
with the antihermitian operator 
\begin{equation}
I_-=\hat{a}^\dagger\sigma_--\hat{a}\sigma_+. 
\end{equation}
Using the Baker Campbell Hausdorff formula we have
\begin{equation}
H=H_F+\frac{\lambda}{2}[I_-,H_F]+\frac{1}{2!}\left(\frac{\lambda}{2}\right)^2[I_-,[I_-,H_F]]+\dots
\end{equation}
The first commutator is worked out explicitly
\begin{equation}
[I_-,H_F]=[\hat{a}^\dagger\sigma_--\hat{a}\sigma_+,H_F].
\end{equation} 
Since $H_F=\omega_c I_+-\omega\hat{N}_a+\xi\sigma_x$ and $[I_-,\hat{N}_a]=0$, with $I_+=\hat{a}^\dagger\sigma_-+\hat{a}\sigma_+$, we only need to evaluate two commutators. The first one gives
\begin{eqnarray}
[I_-,I_+]&=&\nonumber[\hat{a}^\dagger\sigma_--\hat{a}\sigma_+,\hat{a}^\dagger\sigma_-+\hat{a}\sigma_+]\\
&&=\nonumber-2[\hat{a}\sigma_+,\hat{a}^\dagger\sigma_-]\\
&&=\nonumber-2(\hat{a}[\sigma_+,\hat{a}^\dagger\sigma_-]+[\hat{a},\hat{a}^\dagger\sigma_-]\sigma_+)\\
&&=\nonumber-2(\hat{a}\hat{a}^\dagger[\sigma_+,\sigma_-]+[\hat{a},\hat{a}^\dagger]\sigma_-\sigma_+)\\
&&=\nonumber-2(\hat{a}\hat{a}^\dagger\sigma_z+(\mathbb{1}-\sigma_z)/2)\\
&&=\nonumber-2(\hat{a}^\dagger\hat{a}\sigma_z+(\mathbb{1}+\sigma_z)/2)\\
&&=\nonumber-2(\hat{a}^\dagger\hat{a}+(\mathbb{1}+\sigma_z)/2)\sigma_z\\
&&=-2\hat{N}_a\sigma_z,
\end{eqnarray} 
whereas the second one follows as
\begin{eqnarray}
[I_-,\sigma_x]&=&\nonumber[\hat{a}^\dagger\sigma_--\hat{a}\sigma_+,\sigma_-+\sigma_+]\\
&&=\nonumber-\hat{a}[\sigma_+,\sigma_-]+\hat{a}^\dagger[\sigma_-,\sigma_+]\\
&&=\nonumber-(\hat{a}+\hat{a}^\dagger)\sigma_z.
\end{eqnarray}
Upon substitution of these first-order corrections and introduction of the shifted harmonic oscillator operators we get 
the effective Hamiltonian given in equation (\ref{hfin}). 

\end{document}